\documentclass[twocolumn,aps,prl,groupedaddress,showpacs]{revtex4}
\usepackage{epsfig,amssymb,amsmath}

\def\comment#1{}\def\labell#1{\label{#1}}
\begin{document}
%{\scriptsize Eprint: quant-ph/}
\fbox{{\scriptsize Submitted draft.}}
%Title of paper
\title{Quantum metrology} \author{Vittorio Giovannetti,$^1$ Seth
  Lloyd$^{2}$, Lorenzo Maccone$^3$}\affiliation{$^{1}$NEST-INFM \&
  Scuola Normale Superiore, Piazza dei Cavalieri 7, I-56126, Pisa,
  Italy.\\$^{2}$MIT, Research
  Laboratory of Electronics and Dept. of Mechanical Engineering,\\
  77  Massachusetts Avenue, Cambridge, MA 02139, USA.\\
  $^3$ QUIT - Quantum Information Theory Group, Dipartimento di Fisica
  ``A.  Volta'' Universit\`a di Pavia, via A.  Bassi 6, I-27100 Pavia,
  Italy.}
%\date{\today}

\begin{abstract}
  We point out a general framework that encompasses most cases in
  which quantum effects enable an increase in precision when
  estimating a parameter (quantum metrology). The typical quantum
  precision-enhancement is of the order of the square root of the
  number of times the system is sampled. We prove that this is optimal
  and we point out the different strategies (classical and quantum)
  that permit to attain this bound.
\end{abstract}
\maketitle 
When estimating an unknown parameter in a quantum system, we typically
prepare a probe, let it interact with the system, and then measure the
probe. If the physical mechanism which governs the system dynamics is
known, we can deduce the value of the parameter by comparing the input
and the output states of the probe. Since quantum states are rarely
distinguishable with certainty, there usually is an inherent
statistical uncertainty in such estimation. To reduce this
uncertainty, we can use $N$ identical, independent probes, measure
them and average the results.  From the central limit theorem, for
large $N$ the error on the average decreases as $\Delta/\sqrt{N}$,
where $\Delta^2$ is the variance of the measurement results associated
with each probe. Using the same physical resources with the addition
of quantum effects (such as entanglement or squeezing) an even better
precision can often be achieved with a customary $\sqrt{N}$
enhancement, i.e. a scaling of $1/N$~\cite{review}.

In this paper we introduce a theoretical framework that encompasses
all of these strategies and we show that the scaling $1/N$ is the general
lower bound to the estimation error: The only way to further decrease
the error is to reduce $\Delta$, by improving the probe response to
the interaction with the system.  In analogy to quantum
communication~\cite{bennetshor}, different scenarios are possible (see
Fig.~\ref{f:schema}): Either we do not employ quantum effects (CC
strategy), or quantum effects can be used either only in the probe
measurement (CQ strategy), or only in the probe preparation (QC
strategy), or in both stages (QQ strategy).  We will show that the
ultimate precision limit for the CC and CQ strategies is the classical
limit $1/\sqrt{N}$, while the ultimate limit for the QC and QQ
strategies is $1/N$.  This means that, even though entanglement
at the preparation stage is useful to increase the precision, it is
useless at the measurement stage. Hence, the previously proposed
methods for quantum-enhanced parameter estimation can be modified
relinquishing entangled measurements without performance loss.
Moreover, if one is willing to exchange physical resources with
running time, the same precision $1/N$ of the quantum strategy can be
achieved also classically by sequentially applying the transformation
$N$ times on the same probe (multiround protocol, see
Fig.~\ref{f:schema1})~\cite{luis,grover}. We prove optimality also in
this case: No multiround protocol exists that can achieve an error
which scales better than $1/N$.
 
\begin{figure}[hbt]
\begin{center}
\epsfxsize=.75
\hsize\leavevmode\epsffile{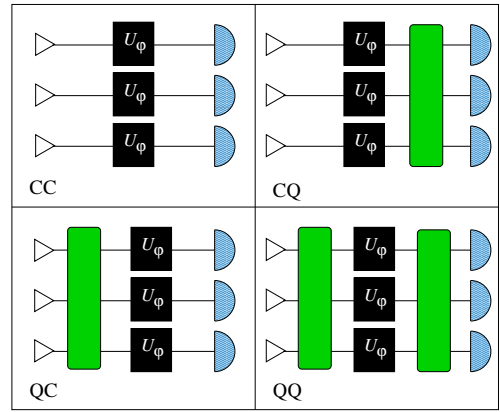}
\end{center}
\vspace{-.5cm}
\caption{Different possible strategies for the estimation of a
  parameter $\varphi$ involving $N$ parallel samplings of a unitary
  operator $U_\varphi$ (black squares). The CC strategy involves
  separable input states and separable measurements (i.e. local
  operations and measurements whose results are communicated
  classically--- LOCC). The CQ strategy involves separable input
  states and general measurement schemes. The QC strategy involves
  general input states (also entangled) and separable measurements.
  The QQ strategy involves general input states and general
  measurement schemes. The triangles on the left represent state
  preparation and the symbols on the right represent measurements. The
  gray boxes represent a unitary operation involving multiple probes
  (Q strategies). } \labell{f:schema}\end{figure}

\begin{figure}[hbt]
\begin{center}
\epsfxsize=.75
\hsize\leavevmode\epsffile{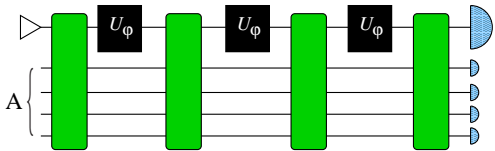}
\end{center}
\vspace{-.5cm}
\caption{Sequential (or multiround) protocol with a single probe. 
  Thanks to the ancillary systems A, this scheme encompasses also
  adaptive techniques where information on $\varphi$ is extracted
  between successive applications of the unitary $U_\varphi$.  As in
  Fig.~\ref{f:schema}, the triangle represents the probe state
  preparation, the black squares represent $U_\varphi$, the symbols on
  the right represent detection, and the gray boxes represent unitary
  operations involving both the probe and the ancillas. }
\labell{f:schema1}\end{figure}

In the case of interferometry, it has long been claimed that, when
using $N$ photons in the interferometer, the Heisenberg limit $1/N$ is
the ultimate bound to precision in phase measurements (whereas
classical strategies only permit to reach the shot noise limit of
$1/\sqrt{N}$~\cite{ou}).  However, the available
proofs~\cite{interf,ou,yurke} are based on an incorrect interpretation
of the time-energy uncertainty relation~\cite{aharonov}, or seem to
lack the necessary generality.  Our analysis clarifies
that indeed the Heisenberg limit is the bound to interferometric
precision. Our bound also applies to quantum phase-estimation
strategies~\cite{qpe}, which are customarily presented as examples of
exponential-speedup algorithms.  In fact, even though a precision
$\sim 2^{-K}$ that scales exponentially with the number $K$ of
employed qubits is achieved, the algorithms require an exponential
number of applications of the unitary $U$ that generates the phase
shift.  Thus, in terms of the number $N\simeq 2^K$ of times that $U$
needs to be employed in the procedure, one finds the same $1/N$
precision scaling of our optimality bound for sequential strategies.

%In Sec.~\ref{s:framework} 
In the following, we first analyze the theoretical framework which
includes most known quantum metrology protocols. We then derive the
bounds to precision in the different scenarios, and show that they are
achievable.  %In Sec.~\ref{s:particular} 
Finally, we show how this relates to the known protocols and how to
generate new protocols.

\paragraph{Theoretical framework:--}\labell{s:framework}
Our goal is to find the most efficient possible way of estimating a
parameter $\varphi$, introduced by the system onto the probe through a
unitary operator $U_\varphi\equiv\exp(-i\varphi H)$, where the
generator $H$ is a known Hermitian operator. If we are allowed to
sample the system $N$ times, we can either use the parallel
configuration of Fig.~\ref{f:schema} where $N$ probes are jointly
employed, or use the sequential configuration of Fig.~\ref{f:schema1}
where a single probe is employed $N$ times (or a combination of these
two strategies).  Notice that the latter configuration is in principle
more powerful than the former. In fact, a sequential strategy can
simulate any other configuration that employs the same number of
$U_\varphi$'s, if we add appropriate ancillas and if we allow the
total running time to increase.

We start by analyzing the parallel strategies.  Taking $|\Psi\rangle$
as the state of the $N$ probes, it will be transformed into
$U_\varphi^{\otimes N}|\Psi\rangle$, where $U_\varphi^{\otimes N}$ is
the unitary transformation generated by $h=\sum_{j=1}^NH_j$ ($H_j$
acting on the $j$ th probe).  In order to take into account the
possibility that $\varphi$ can be estimated through a general (biased
or unbiased) estimator, it is convenient to use the error
estimate~\cite{holevo,caves}
\begin{eqnarray}
\delta\varphi\equiv\left\langle\left(\varphi_{est}/\left|\frac
{\partial\langle\varphi_{est}\rangle}
{\partial\varphi}\right|-\varphi\right)^2\right\rangle
\;\labell{defdelta},
\end{eqnarray}
where $\varphi_{est}$ is the estimator employed and where the brackets
$\langle\;\rangle$ denote statistical averaging (the purpose of the
derivative $\partial\langle\varphi_{est}\rangle/\partial\varphi$ is
simply to express both $\varphi_{est}$ and $\varphi$ in the same
``units'').  Whatever is the measurement scheme employed, the error
$\delta\varphi$ is bounded by the generalized uncertainty
relation~\cite{caves},
\begin{eqnarray}
\delta\varphi\:\Delta h\geqslant 1/({2\sqrt{\nu}})
\;\labell{caves},
\end{eqnarray}
where $(\Delta h)^2=\langle h^2\rangle-\langle h\rangle^2$ is the
variance of $h$ on the input state $|\Psi\rangle$ of the $N$ probes,
and $\nu$ is the number of times the estimation is repeated.
Equation~(\ref{caves}) derives from the Cramer-Rao bound and is
asymptotically achievable in the limit of large $\nu$. It implies that
the minimum error $\delta\varphi$ is obtained when $\Delta h$ is
maximum.  If $|\Psi\rangle$ is separable (CC and CQ strategies),
$\Delta h=[\sum_j\Delta^2H_j]^{1/2}$ where $\Delta^2H_j$ is the
variance of $H_j$ on the state of the $j$th probe. Hence the maximum
$\Delta h$ is achieved by preparing each probe in a state having
maximum spread for $H_j$, i.e. the equally weighted superpositions of
the eigenvectors $|\lambda_{M}\rangle$ and $|\lambda_{m}\rangle$ of
$H_j$ corresponding respectively to the maximum and minimum
eigenvalues $\lambda_{M}$ and $\lambda_{m}$.  Thus, $\Delta
h\leqslant\sqrt{N}(\lambda_{M}-\lambda_{m})/2$, which, through
Eq.~(\ref{caves}), gives an optimal CC- and CQ-error of
\begin{eqnarray}
\delta\varphi\geqslant 1/[{\sqrt{\nu\:N}(\lambda_{M}-\lambda_{m})}]
\;\labell{cccq}.
\end{eqnarray}
This bound can be attained, for instance, by Ramsey interferometry,
i.e.  by preparing all the probes in the state
$(|\lambda_{M}\rangle+|\lambda_{m}\rangle)/\sqrt{2}$, and by measuring
the probability that each probe remains unchanged at the output.  Even
though Ramsey interferometry does not employ entangled measurements,
these are accounted for in the derivation of Eq.~(\ref{caves}),
see~\cite{caves}.  This proves that entangled measurements are not
necessary to achieve~(\ref{cccq}): The CC strategy is as accurate as
the CQ strategy~\cite{nota}. 

On the other hand, if $|\Psi\rangle$ can be entangled (QC and QQ strategies), the
maximum $\Delta h$ corresponds to a $|\Psi\rangle$ which is an equally
weighted superposition of the eigenvectors relative to the maximum and
minimum eigenvalues of the global generator $h$, i.e.
$N\lambda_{M}$ and $N\lambda_{m}$. Thus, $\Delta h\leqslant
N(\lambda_{M}-\lambda_{m})/2$, which, through Eq.~(\ref{caves}),
gives an optimal QC- and QQ-error of
\begin{eqnarray}
\delta\varphi\geqslant 1/[{\sqrt{\nu\:}N(\lambda_{M}-\lambda_{m})}]
\;\labell{qcqq},
\end{eqnarray}
with a $\sqrt{N}$ improvement over Eq.~(\ref{cccq}). Notice that the
derivation still applies if $|\Psi\rangle$ includes some external
ancillas in addition to the probes, so that Eq.~(\ref{qcqq}) accounts
also for those detection strategies where half of an entangled state
is fed into the system and a joint measurement is
performed~\cite{paolop}.  Also the bound~(\ref{qcqq}) is attainable:
Use the following entangled state of $N$ probes
\begin{eqnarray}
|\Psi\rangle=\frac 1{\sqrt{2}}
\Big
(|\lambda_{m}\rangle_1\cdots|\lambda_{m}\rangle_N
+|\lambda_{M}\rangle_1\cdots|\lambda_{M}\rangle_N
\Big)
\;\labell{stent},
\end{eqnarray}
and estimate $\varphi$ by measuring the observable
$X\equiv|\lambda_{m}\rangle\langle\lambda_{M}|+|\lambda_{M}\rangle\langle\lambda_{m}|$
separately on each probe at the output (an LOCC strategy).  Since
$\langle X^{\otimes
  N}\rangle_{out}=\cos[N\varphi(\lambda_{M}-\lambda_{m})]$ and the
variance $\Delta X^{\otimes
  N}=\left|\sin[N\varphi(\lambda_{M}-\lambda_{m})]\right|$, after
repeating $\nu$ times the experiment the error on $\varphi$ can be
obtained easily from error propagation as
 \begin{eqnarray}
\delta\varphi= \frac{1}{\sqrt{\nu}}\;\Delta X^{\otimes N}/\left|\frac{\partial\langle X^{\otimes
N}\rangle}{\partial\varphi}\right|=\frac 1{\sqrt{\nu}\:N(\lambda_{M}-\lambda_{m})}
\;\labell{prog}.
\end{eqnarray}
This procedure attains the bound~(\ref{qcqq}) and again only employs
separable measurements.  This proves that entangled measurements are
not necessary to achieve~(\ref{qcqq}): The QC strategy is as accurate
as the QQ strategy.  It has been pointed out that the above
$\delta\varphi$ refers only to the determination of the last
significant digits of $\varphi$~\cite{burgh}. If one wants to
determine all digits of $\varphi$, the procedure must be changed, but
the $1/N$ scaling persists. For example, one can use a single probe
$\nu$ times to recover the first decimal digit of $\varphi/2\pi$.
Then, one can entangle $10$ probes and determine the second decimal
digit, still with $\nu$ repetitions.  Iterating, the $j$th decimal
digit will need $10^j$ entangled probes.  Thus, the total number of
probes (employed $\nu$ times) to recover $l$ decimal digits is
$\sum_{j=0}^l10^l=(10^{l+1}-1)/9$: Almost all the probes ({roughly}
a fixed fraction $b-1/b$, when using $b$-ary notation) are employed to
determine the last digit only.

Instead of a parallel strategy on $N$ probes, one can employ a
sequential strategy on a single probe. In this case the generator $h$
in Eq.~(\ref{caves}) must be modified: Instead of referring to the
unitary $U_\varphi^{\otimes N}$ acting on $N$ probes, it now refers to
a unitary $W_\varphi$ which contains $N$ applications of $U_\varphi$
on a single probe, i.e.
$W_\varphi=V_NU_{\varphi}V_{N-1}U_{\varphi}\cdots V_1U_\varphi V_0$.
Here the $V_j$'s are arbitrary unitary operators acting on the probe
and, eventually, on ancillary systems that can be used in adaptive
strategies to extract information during the estimation process (i.e.
the gray boxes of Fig.~\ref{f:schema1}). In this case, the generator
of $W_\varphi$ is $h\equiv i({\partial W_\varphi}/{\partial\varphi})
W_\varphi^{\dag}\equiv\sum_{j=1}^N{H}'_j(\varphi)$, where
$H'_j(\varphi)\equiv V_jU_\varphi\cdots V_1U_\varphi V_0\:H \:V_0^\dag
U_\varphi^\dag V_1^\dag\cdots U_\varphi^\dag V_j^\dag$ ($H$ being the
generator of $U_\varphi$). Since all the $H'_j$ have the same
spectrum as $H$, then the maximum eigenvalue of $h$ is upper bounded
by $N\lambda_{M}$, while the minimum eigenvalue of $h$ is lower
bounded by $N\lambda_{m}$. Hence, $\Delta h \leqslant N (\lambda_M -
\lambda_m)/2$, and Eq.~(\ref{caves}) in this case implies
\begin{eqnarray}
\delta\varphi\geqslant 1/[{\sqrt{\nu\:}N(\lambda_{M}-\lambda_{m})}]
\;\labell{qcqq1}.
\end{eqnarray}
It is identical to the QC-QQ bound of Eq.~(\ref{qcqq}), even though it
refers to a different physical situation. This bound is again
achievable through Ramsey interferometry, by preparing the single
probe in the state
$(|\lambda_{M}\rangle+|\lambda_{m}\rangle)/\sqrt{2}$, applying to it
the transformation $U^N_\varphi$, and measuring the probability that
it remains unchanged.  Notice that the same analysis is valid also when
the $U_\varphi$'s are applied to more than one probe, i.e. for the 
strategies which are intermediate between the parallel and the
sequential one.

The QC and QQ protocols may seem less appealing than the multiround
protocol since they require entanglement among the $N$ probes to
achieve the same sensitivity. However, their parallelizable structure
entails that their running time may be $N$ times smaller than the
running time of the (necessarily sequential) multiround protocol. This
is one of the instances frequently encountered in quantum computation
where entanglement can convert spatial resources into temporal
resources.  

The above analysis illustrates how entanglement permits the full
exploitation of the Hilbert space of $N$ probes, granting access to
`high-resolution states' such as the one given in Eq.~(\ref{stent}).
In repeating the process $\nu$ times, we can then achieve a precision
that scales as $1/(N\sqrt{\nu})$ for large $\nu$.  This is a purely
quantum effect. In fact, in a classical setting there is no advantage
in grouping the measurements into $\nu$ groups of $N$: The error will
invariably scale as $1/\sqrt{N\:\nu}$, i.e.  as the inverse of the
square root of the total number of measurements.

\paragraph{Quantum metrology protocols:--}\labell{s:particular}
Most quantum metrology protocols can be analyzed under the
theoretical framework outlined above.  In particular, interferometric
strategies can be accounted for by identifying $N$ with the total
number of passes of the employed photons through the interferometer,
and the generator $h$ with the electromagnetic field Hamiltonian.
Here, the $1/N$ scaling of the optimal precision coincides with the
Heisenberg limit, and it is well known that such a limit can be
attained through entangled or squeezed light at the input ports of the
interferometer (e.g.  see~\cite{yurke}), or through multiround
protocols~\cite{luis}. The quantum-positioning and
clock-synchronization protocol~\cite{qcs} is an example of
interferometric strategy where a $1/N$ scaling in the precision of
localization is obtained using frequency-entangled or number-squeezed
photons in a parallel configuration. The same results can be achieved
also in a sequential configuration by bouncing back and forth a single
photon~\cite{burgh}.

Our framework encompasses many other estimation strategies. An example
is the quantum-frequency-standards procedure~\cite{interf,freqstd},
where the collective behavior of entangled atoms is used to enhance
the precision of frequency measurements. It can be analyzed in our
framework by identifying $H_j$ with the two-level Hamiltonian of each
probe atom.  In this context, it is interesting to note that, in
agreement with the equivalence between the QC and QQ strategies, one
can achieve the upper bound~(\ref{qcqq}) measuring separately the
population of each atom~\cite{burgh}, without resorting to the
entangled measurement of the original proposals.

Using our framework, it is also possible to design new quantum
metrology protocols. For example, by entangling $N$ particles in
momentum, we can design a strategy to obtain a better precision in the
measurement of their average position from position measurements on
the single particle (notice that in Ref.~\cite{qcs} the average
position was deduced from time-of-arrival measurements and not from
position measurements).

Even though we assumed that the operator $H$ (i.e. the generator of the
unitary $U_\varphi$) is known, the bounds we derived are valid also if
$H$ is unknown. However, in this case it is not granted that such
bounds are achievable: All our `achievability' protocols require the
knowledge of the eigenstates of $H$. Nonetheless, at least in the case
of the reference-frames-transmission (a procedure to employ $N$ spins
in transmitting a reference frame to a distant party, in which $H$ is
not known because it is the object to be estimated), a protocol
achieving a scaling of $1/N$ has been recently
proposed~\cite{refframes}.

\paragraph{Conclusions:--}\labell{s:conclusion}
State preparation is the primary factor in boosting the precision of
the parameter estimation, while entangled measurements are never
necessary. A $\sqrt{N}$ precision-enhancement over what can be
attained with a classical parallel strategy is typically obtained by
using an input state that is entangled on a basis of eigenstates of
$H$ (the generator of the unitary $U_\varphi$), and by measuring a set
of projectors on a basis dual to that.  Schematically: 1)~entangle $N$
probes on the basis of eigenstates of $H$; 2)~let the probes interact
with the system; 3)~measure on a dual basis. Result: a $\sqrt{N}$
precision enhancement.  This is clearly related to the fact that
entangled states can evolve faster than unentangled configurations
employing the same resources~\cite{speedlimit}.  Alternatively, a
multiround protocol can achieve the same optimal precision at the
expense of a larger running time.

\par S. L. acknowledges financial support by ARDA, DARPA,
ARO, AFOSR, NSF, and CMI; V. G. was in part supported by the EC under
contract IST-SQUBIT2; L. M. acknowledges financial support by MIUR
through ``Cofinanziamento 2003'' and by EC through ATESIT (Contract
No.~IST-2000-29681).


\begin{references}
\bibitem{nota} Equation~(\ref{cccq}) applies also when the probes are
  prepared into a separable mixture,
  $\sigma=\sum_kp_k\:\rho^{(1)}_k\otimes\cdots\otimes\rho^{(N)}_k$.
  The best experimental situation is when the information on the
  random variable $k$ is accessible. In this case, the global state of
  $\nu$ experimental runs can be written as
  $\bigotimes_k\;[\rho_k^{(1)}\cdots\rho_k^{(N)}]^{\otimes n_k}$,
  where $n_k=p_k \nu$ is the number of times in which the $k$th
  component is employed. Starting from this tensor product form, we
  can employ the above derivation to obtain again the
  bound~(\ref{cccq}), which hence applies also to separable mixed
  states.
\end{references}
\end{document}